\documentclass[preprint,aps,tightenlines,floatfix,showpacs]{revtex4}
\usepackage{graphics}
\usepackage{amsmath}
\usepackage{bm}
\begin{document}
\preprint{LA-UR-02-987}
\title{Predictions of total and total reaction cross sections for
nucleon-nucleus scattering up to 300 MeV}
\author{K. Amos}
\email{amos@physics.unimelb.edu.au}
\affiliation{School of Physics, The University of Melbourne, Victoria
3010, Australia}
\author{S. Karataglidis} 
\email{stevenk@lanl.gov}
\affiliation{Theoretical Division, Los Alamos National Laboratory, Los
Alamos, New Mexico, 87545}
\author{P. K. Deb}
\email{deb@physics.unimelb.edu.au}
\affiliation{School of Physics, The University of Melbourne, Victoria,
3010, Australia}
\date{\today}
\begin{abstract}
Total reaction cross sections are predicted for nucleons scattering
from various nuclei. Projectile energies to 300 MeV are considered.
So also are mass variations of those cross sections at selected
energies.  All predictions have been obtained from coordinate space
optical potentials formed by full folding effective two-nucleon ($NN$)
interactions with one body density matrix elements (OBDME) of the
nuclear ground states.  Good comparisons with data result when
effective $NN$ interactions defined by medium modification of free
$NN$ $t$ matrices are used.  Coupled with analyses of differential
cross sections, these results are sensitive to details of the model
ground states used to describe nuclei.
\end{abstract}
\maketitle

\section{Introduction}
Reaction cross sections from the scattering of nucleons by nuclei
(stable and radioactive) are required in a number of fields of study;
some being of quite current interest \cite{De01}. An example is the
transmutation of long lived radioactive waste into shorter lived
products using accelerator driven systems (ADS). These systems are
being designed in the US, Europe, and Japan with the added objective
of providing an intense neutron source to a sub-critical reactor as a
new means of energy production and for which nucleon-nucleus reaction
cross sections are required as well.  The technology takes advantage
of spallation reactions \cite{Wl00} within a thick high-$Z$ target
(such as Pb or Bi), where an intermediate energy proton beam induces
nuclear reactions. The secondary nuclear products, particularly lower
energy neutrons and protons \cite{Le99}, in turn induce further
nuclear reactions in a cascade process. The total reaction cross
sections of nucleon-nucleus scattering plays a particularly important
role since the secondary particle production cross sections are
directly proportional to them. Also they are inputs to intra-nuclear
cascade simulations that guide ADS design.  As well, nucleon-nucleus
($NA$) cross section values at energies to 300 MeV or more are needed
not only to specify important quantities of relevance to proton and
neutron radiation therapy \cite{Jo01}, but also as they are key
information in assessing radiation protection for patients.

In basic science, these total reaction cross sections are important
ingredients to a number of problems in astrophysics, such as
nucleosynthesis in the early universe and for aspects of stellar
evolution, especially as the density distribution of neutrons in
nuclei are far less well known than that of protons.  Recently a link
has been made between the neutron distribution in heavy nuclei, such
as $^{208}$Pb, and properties of neutron stars \cite{Br00,Ho01} so
heightening the need for analyses to ascertain the optimal neutron
distributions in such nuclei.  Analyses of proton elastic scattering
angular distributions particularly for 200~MeV protons seem to be one
such method \cite{Ka02}.  Considering the integral observables of both
proton and neutron scattering from a given nucleus is another and
certainly it will give direct information on the neutron rms radius; a
property sought in new parity-violating electron scattering
experiments \cite{Ka02,Br00}.

However, most $NA$ reaction cross sections cannot be, have not been,
or are unlikely to be, measured.  Thus a reliable method for their
prediction is required.  The usual vehicle for specifying these $NA$
total reaction cross sections has been the $NA$ optical potential; a
potential most commonly taken as a local parametrized function,
usually of Woods-Saxon (WS) type.  However, it has long been known
that the optical potential must be nonlocal and markedly so, although
it has been assumed also that such nonlocality can be accounted by the
energy dependence of the customary (phenomenological) models
\cite{Am00}.  Of more concern is that the phenomenological approach is
not truly predictive.  The parameter values chosen, while they may be
set from a global survey of data analyses, are subject to considerable
uncertainties and ambiguities. This is especially true for the optical
potentials for nucleon scattering from $0p$-shell nuclei where no such
global approach is valid.

We consider a predictive theory of $NA$ scattering to be one that is
``direct'' in that all quantities required are defined \textit{a
priori}.  Thus each result must come from just one run of relevant
codes, and there should be no post evaluation adjustments save for
allowance of known \textit{a priori} uncertainties in the
specification of the input information.

With the nucleus viewed as a system of $A$ nucleons, $NA$ scattering
is determined to first order by an optical potential formed by
folding, with a suitable specification of the ground state density of
the target, appropriate interactions of that projectile with each and
every nucleon within the nucleus.  Over the past decade or more, such
microscopic approaches defining the $NA$ optical potential have been
quite successful in predicting elastic scattering data (differential
cross sections in particular).  Studies forming the optical potential
in both momentum and coordinate space have been made with success
\cite{Am00}.  In the coordinate space approach, and for analyses that
are based upon the DWBA programs of Raynal \cite{Ra91,Ra98}, the
projectile-target nucleon interaction takes the form of a complex,
energy and density dependent, effective $NN$ interaction.  Of those
programs, DWBA98 has been used to evaluate all of the cross sections
shown later herein.  Appropriate effective interactions can and have
been defined that, upon folding with good structure wave functions of
nuclei, give credible optical potentials. Using those optical
potentials, differential cross sections and spin observables such as
the analyzing powers for proton scattering at many energies in the
range 40 to 800~MeV (65 and 200 MeV in particular) and from diverse
targets ranging from $^3$He to $^{238}$U have been predicted and found
to have excellent agreement with data.  Moreover, and very recently
\cite{Ka02}, it has been shown that analyses of differential cross
sections of proton elastic scattering can select between alternative
model predictions of the neutron rms radius in $^{208}$Pb; such are
sensitive also to the surface distribution of its neutron matter.

The microscopically formed optical potentials are complex and energy
dependent from the like properties of the effective $NN$ interaction
\cite{Am00}. Such properties arise from mapping the effective
interactions to $NN$ $g$ matrices that are solutions of the
Bruckner-Bethe-Goldstone (BBG) equations for nuclear matter.  The BBG
equations carry medium modification of the $NN$ scattering due to
Pauli blocking and to a background mean field.  Details of the
effective interactions, of the folding process that gives the
(nonlocal) optical potential, and of the successful predictions found
therefrom of differential cross sections and analyzing powers from the
scattering of protons at diverse energies and from diverse targets,
are given in the literature \cite{Am00}.

\section{Elements of the optical potential for nucleon scattering}

Formally, the nonlocal optical potentials from a first order folding
model can be written
\begin{eqnarray}
U(\bm{r}_1,\bm{r}_2; E) & = & \sum_n \zeta_n \left\{ \delta(\bm{r}_1
- \bm{r}_2) \int \varphi^*_n(\bm{s})\, v_D(\bm{r}_{1s})\,
\varphi_n(\bm{s}) \, d\bm{s} + \varphi^*_n(\bm{r}_1)\,
v_{Ex}(\bm{r}_{12})\, \varphi_n(\bm{r}_2) \right\} \nonumber\\
& \Rightarrow & U_D(\bm{r}_1; E) \delta(\bm{r}_1 - \bm{r}_2)
+ U_{Ex}(\bm{r}_1,\bm{r}_2; E)\, ,
\label{NonSE}
\end{eqnarray}
where $v_D$, $v_{Ex}$ are combinations of the components of the
effective $NN$ interactions, $\zeta_n$ are ground state nucleon shell
occupancies (more generally they are the ground state OBDME), and
$\varphi_n(\bm{x})$ are nucleon bound state wave functions; denoted SP
functions hereafter.  All details and the prescription of solution of
the associated nonlocal Schr\"odinger equations are given in the
review \cite{Am00}.

The results to be discussed have been found by solving the actual
nonlocal Schr\"odinger equations defined with potentials as given
(formally) by Eq.~(\ref{NonSE}).  Two formulations of those optical
potentials have been used.  They and the results are identified by the
appellations, $g$- and $t$- folding, according that the effective $NN$
interactions have been defined by their mapping to the BBG $g$
matrices or to the basic free $NN$ scattering $t$ matrices
respectively. The latter are solutions of Lippmann-Schwinger (LS)
equations.  In both cases, the driving $NN$ interaction has been the
Bonn-B $NN$ potential \cite{Ma87}.

From practical necessity the model descriptions of nuclei in the mass
range 3 to 238 vary in complexity.  With the light mass nuclei ($A \le
12$ for example), quite large and complete shell model spaces with
potentials either fitted or formed as $G$-matrix elements have been
made \cite{Zh95,Na98,Ka00}.  While large space model studies of
heavier nuclei are being sought, the dimensions of the problem
preclude our use of all but $0\hbar\omega$ shell model specifications
for most heavier nuclei.  Indeed, for targets heavier than mass 90 we
have used an even simpler, packed shell, definition of their ground
states.  Nevertheless with such model prescriptions and using harmonic
oscillator (HO) SP functions with oscillator energies selected
according to an $A^{-\frac{1}{3}}$ rule, very good predictions of the
scattering of 65 and 200 MeV protons have been obtained for all but
the light mass nuclei \cite{Am00}.  However, $^{208}$Pb and $^{40}$Ca
are special cases. Recently \cite{Br00,Ka02}, a Skyrme-Hartree-Fock
(SHF) model of those nuclei was made and OBDME required in our
folding procedure were extracted.  The associated density
distributions vary noticeably from that given by the HO (packed) shell
model and, not surprisingly, so do proton differential cross sections.

For nuclei with $A \le 12$ typically, better spectroscopy is needed.
So also are more realistic matter distributions for studies with light
mass exotic nuclei, such as of radioactive beam scattering from
hydrogen.  For example, the reaction cross section for $40.9A$ MeV
$^6$He scattering from hydrogen \cite{La01} varies from 350~mb, found
when $^6$He has a neutron skin as expected with a standard shell model
description, to 406~mb when that distribution is extended further to
be classified as a halo by choosing valence neutron SP functions
consistent with the single neutron separation energy in $^6$He. The
measured value is $409 \pm 21$~mb \cite{De00,La01}.

\section{Phase shifts, S-matrices, and observables}
Irrespective of the means used to define $NA$ optical potentials, the
objective is to define the $S$ matrix, or equivalently the (complex)
phase shifts $\delta^{\pm}_l(k)$, where the superscripts identify the
values $j= l\pm 1/2$.  These relate by
\begin{equation}
S^{\pm}_l(k) = e^{2i\delta_l^{\pm}(k)} = \eta^{\pm}_l(k)
e^{2i\Re\left[ \delta^{\pm}_l(k) \right]}
\end{equation}
where
\begin{equation}
\eta^{\pm}_l(k) = \left| S^{\pm}_l(k) \right| =
e^{-2\Im\left[ \delta^{\pm}_l(k) \right]} \; .
\end{equation}
With $E \propto k^2$, the elastic, reaction (absorption), and total
cross sections respectively then are given by
\begin{eqnarray*}
\sigma_{\text{el}}(E) & = & \frac{\pi}{k^2} \sum^{\infty}_{l = 0} \left\{
\left( l + 1 \right) \left| S^+_l(k) - 1 \right|^2 + l \left|
S^-_l(k) - 1 \right|^2 \right\} \; , \\
\sigma_{\text{R}}(E) & = & \frac{\pi}{k^2} \sum^{\infty}_{l = 0} \left\{
\left( l + 1 \right) \left[ 1 - \eta^+_l(k)^2 \right] + l \left[ 1 -
\eta^-_l(k)^2 \right] \right\} \; ,
\end{eqnarray*}
and
\begin{eqnarray*}
\sigma_{\text{TOT}}(E) & = & \sigma_{\text{el}}(E) +
\sigma_{\text{R}}(E) \\
& = & \frac{2\pi}{k^2} \sum^{\infty}_{l = 0} \left\{ \left( l + 1
\right) \left[ 1 - \eta^+_l(k)\cos\left( 2\Re\left[ \delta^+_l(k)
\right] \right) \right] + l \left[ 1 - \eta^-_l(k) \cos\left(
2\Re\left[ \delta^-_l(k) \right] \right) \right] \right\} \;
. \\
\end{eqnarray*}

The scattering amplitudes are then $2 \times 2$ matrices in the nucleon
spin space and have the form,
\begin{equation}
f(\theta) = g(\theta) + h(\theta) \bm{\sigma} \cdot \hat{\bm{n}}\; .
\label{spinamp}
\end{equation}
where
\begin{eqnarray}
g(\theta) & = & \frac{1}{k} \sum_{l=0} \left\{ \left( l + 1 \right)
\left[ S^+_l(k) - 1 \right] + l \left[ S^-_l(k) - 1 \right] \right\}
P_l(\theta) \; , \nonumber \\
h(\theta) & = & \frac{1}{ik} \sum_{l=1} \left[ S^+_l(k) - S^-_l(k)
\right] P_l^1(\theta) \; .
\end{eqnarray}
In terms of these (complex) amplitudes, the (elastic scattering)
differential cross section is defined by
\begin{equation}
\frac{d\sigma}{d\Omega} = \left| g(\theta) \right|^2 + \left|
h(\theta) \right|^2 \; ,
\end{equation}
and the analyzing power $A_y(\theta)$ by
\begin{equation}
A_y(\theta) = \frac{ 2 \Re\left[ g^*(\theta)h(\theta) \right] }{
d\sigma/d\Omega }
\end{equation}

\section{Results of calculations}

All results we show have been evaluated using the DWBA98 program
\cite{Ra98}, input to which are density dependent and complex
effective $NN$ interactions having central, two-nucleon tensor, and
two-nucleon spin-orbit components.  The effective interactions we use
have been generated for energies from 10 MeV to over 300 MeV in 10 MeV
steps by an accurate mapping to $NN$ $t$- and $g$ matrices found by
solutions of the LS and BBG equations respectively and based usually
upon the Bonn $NN$ potentials.  Details are given in the review
\cite{Am00}.

Other input to DWBA98 are the ground state occupancies (or OBDME) and
the associated SP functions.  The SP functions used in most of the
calculations for nuclei of mass 20 and above at best come from a
$0\hbar\omega$ shell model which has been adopted to describe their
ground state occupancies.  For the lighter mass nuclei considered,
larger shell model spaces were used to define their ground states, and
in some cases the interaction potentials defined as $G$ matrix
elements of a realistic interaction \cite{Zh95}.  In shell model
studies using those $G$-matrix elements, the oscillator energy
($\hbar\omega$) for the SP functions also is specified.  As stated
earlier, the cases of $^{208}$Pb and $^{40}$Ca are special in that we
have used structure information taken from recent SHF studies
\cite{Br00,Ka02}.

\subsection{Energy variation of proton total reaction cross sections}

In this subsection we present our predictions of the total reaction
cross sections for proton scattering up to 300~MeV for diverse nuclei,
ranging in mass from $^6$Li to $^{238}$U.  In all cases, at least two
calculations were made. The first of these used the effective
interaction defined from the $t$ matrices of the Bonn-B interaction
while with the second, that built upon the associated $g$ matrices was
used.  Comparison of the results of each pair of calculations
demonstrates the effects in predictions due to the medium modification
to the free NN interaction that define the $g$ matrices.  The ensuing
$t$- and $g$-folding results are portrayed in the figures by the
dashed and solid curves respectively.

As noted, the structure models of the light mass nuclei involve
diverse complete $N\hbar\omega$ bases.  For the lightest, $^6$Li, a
complete $(0+2+4)\hbar\omega$ model of structure has been used, while
for $^9$Be and $^{12}$C the OBDME have been defined from complete
$(0+2)\hbar\omega$ shell model calculations~\cite{Ka95}.  In addition
we have calculated the reaction cross sections from $^{118}$Sn and
$^{159}$Tb allowing the outer (neutron) shell to have a smaller
(15-20\%) harmonic oscillator energy.  By that means, the neutron
surface of each is slightly more extended than with the base (packed
shell) model forms; a very simple allowance for any effect of ground
state deformations.  This idea for varied surface SP functions within
an HO model has been used in the guise of a two frequency shell model
\cite{Co01}.  Our wave functions are not so well determined of course.

\begin{figure}
\scalebox{0.8}{\includegraphics{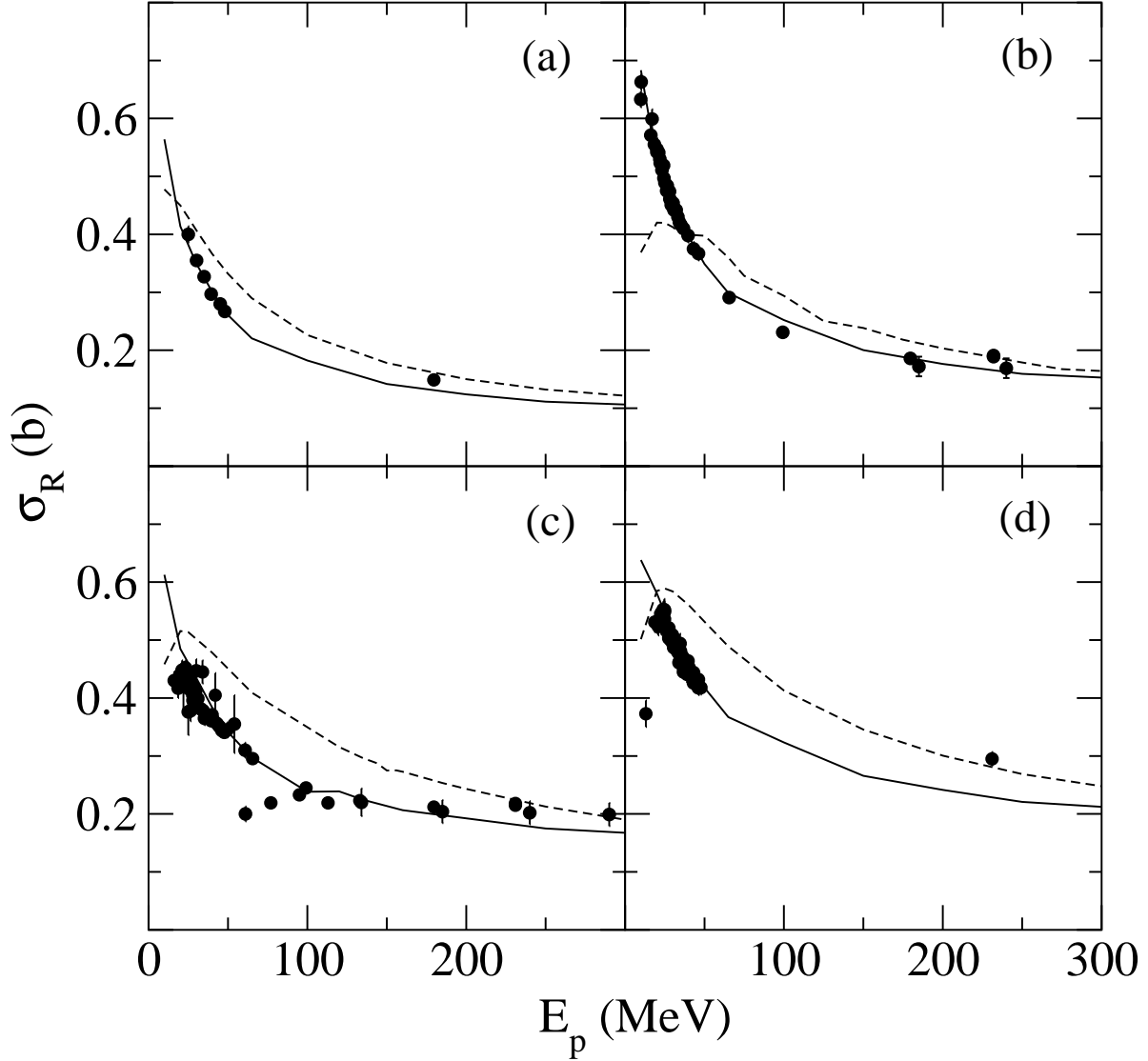}}
\caption{\label{react-Li-Be-C-O} Energy dependencies of $\sigma_R$ for
proton scattering from (a) $^6$Li, (b) $^9$Be, (c) $^{12}$C, and (d)
$^{16}$O. The solid curves are predictions made using $g$-folding
optical model calculations while those found using $t$-folding optical
potentials are displayed by the dashed curves.}
\end{figure}

The results for scattering from $^6$Li, $^9$Be, $^{12}$C and from
$^{16}$O are displayed respectively in segments (a), (b), (c) and (d)
of Fig.~\ref{react-Li-Be-C-O}.  The experimental data shown therein
were taken from the references listed against those nuclei in the data
table, Table~\ref{exptab}; which also lists the sources of the data
that are shown in the figures to follow.
\begin{table}
\begin{ruledtabular}
\caption{\label{exptab} Data source table for proton and neutron reaction data used}
\begin{tabular}{cl}
 Nucleus   & Proton references (in year order)\\
\hline
$^6$Li     & \cite{Jo61}, \cite{Ca75}\\
$^9$Be     & 
\cite{Mi54}, \cite{Jo61}, \cite{Wi63}, \cite{Ki66},
\cite{Re72}, \cite{Mo73}, \cite{Mc74}, \cite{Sl75},
\cite{In99}\\
$^{12}$C   & 
\cite{Mi54}, \cite{Ca54}, \cite{Bu59}, \cite{Go59},
\cite{Me60}, \cite{Jo61}, \cite{Go62},
\cite{Gi64}, \cite{Ma64}, \cite{Ki66}, \cite{Me71},
\cite{Re72}, \cite{Mc74}, \cite{Sl75},
\cite{In99}\\
$^{16}$O   & \cite{Ch67}, \cite{Re72}, \cite{Ca75}\\
$^{19}$F   & \cite{Ch67}, \cite{Ca75}\\
$^{27}$Al  & 
\cite{Mi54}, \cite{Ca54}, \cite{Go59},
\cite{Me60}, \cite{Me60a}, \cite{Jo61}, \cite{Go62}, \cite{Wi63},
\cite{Ma64}, \cite{Be65}, \cite{Po65}, \cite{Ki66}, \cite{Me71},
\cite{Re72}, \cite{Mc74}\\
$^{40}$Ca  &
\cite{Jo61}, \cite{Tu64}, \cite{Ki66}, \cite{Di70}, \cite{Ca75}, \cite{In99}\\
$^{63}$Cu  &
\cite{Ca54}, \cite{Al61}, \cite{Go62}, \cite{Wi63}, \cite{Ma64a},
\cite{Ma64}, \cite{Be65}, \cite{Po65}, \cite{Ki66}, \cite{Di67},
\cite{Ho68}, \cite{Re72}\\
$^{90}$Zr  & \cite{Wi63}, \cite{Ki66}, \cite{Di67}, \cite{Me71}\\
$^{118}$Sn & \cite{Go59}, \cite{Me60}, \cite{Wi63}, \cite{Ki66},
\cite{Di67}, \cite{Re72}, \cite{Ca95}, \cite{In99}\\
$^{140}$Ce & \cite{Da77}\\
$^{159}$Tb & \cite{Ki66}, \cite{Ab79}\\
$^{181}$Ta & \cite{Wi63}, \cite{Ki66}, \cite{Ab79}\\
$^{197}$Au & \cite{Jo61}, \cite{Wi63}, \cite{Ma64}, 
\cite{Be65}, \cite{Ki66}, \cite{Ho68}, \cite{Ab79}\\
$^{208}$Pb & \cite{Ca54}, \cite{Go59}, \cite{Me60}, \cite{Go62},
\cite{Tu64}, \cite{Po65}, \cite{Ki66}, \cite{Me71}, \cite{Re72},
\cite{Mo73}, \cite{Ca75}, \cite{In99}\\ 
$^{238}$U  & \cite{Mi54}, \cite{Ki66}\\
\end{tabular}
\end{ruledtabular}
\end{table}
The data from the four lightest mass nuclei are well reproduced by
$g$-folding calculations made with large space shell model structure
(solid curve), but they are not with $t$-folding calculations (dashed
curve).  However, large space structure calculations are necessary if
one is to describe the physics even adequately.  As indicated above,
for $^6$Li, such structure was found from a shell model
calculation \cite{Ka97} made using a compete $(0+2+4)\hbar\omega$
space while those for $^9$Be, $^{12}$C and $^{16}$O were made using
complete $(0+2)\hbar\omega$ models.  Results found using the simpler
$0\hbar\omega$ structure model of Cohen and Kurath \cite{Co65} within
$g$-folding, underestimate the data at most energies, and for $^9$Be
particularly. This reflects the too compressed density profile for the
nuclei given by the simpler model.
Results are displayed for proton energies from 10~MeV. Although
experimental data exist to lower energies in these cases, we do not
consider the first order folding prescription for the optical
potential to be appropriate in the low energy regime of scattering
from these nuclei.  Up to and over 20~MeV excitation, their spectra
have numerous distinguishable states. Excitations to regions of low
level density are not taken into account forming the optical
potentials.  We have confidence in the optical potentials when the
input energy coincides to excitations to regions of high level density
and where particle emission is feasible.

The $^{12}$C results in particular are worth comment.  For this
nucleus, as with the others, the reaction cross sections obtained from
those $g$-folding calculations for $^{12}$C are in very good agreement
with the experimental data up to 300 MeV.  Most evidently, the medium
effects differentiating the $g$- from the $t$- matrices used in the
folding scheme defining the optical potentials are required for
predictions to match observation.  The $t$-folding model overestimates
the data by 20-40\% within the energy regime below 200 MeV.  But some
data, notably at 61~MeV \cite{Me60} and at 77~MeV \cite{Go62} MeV with
$^{12}$C, are in disagreement with both calculated results.

\begin{figure}
\scalebox{0.8}{\includegraphics{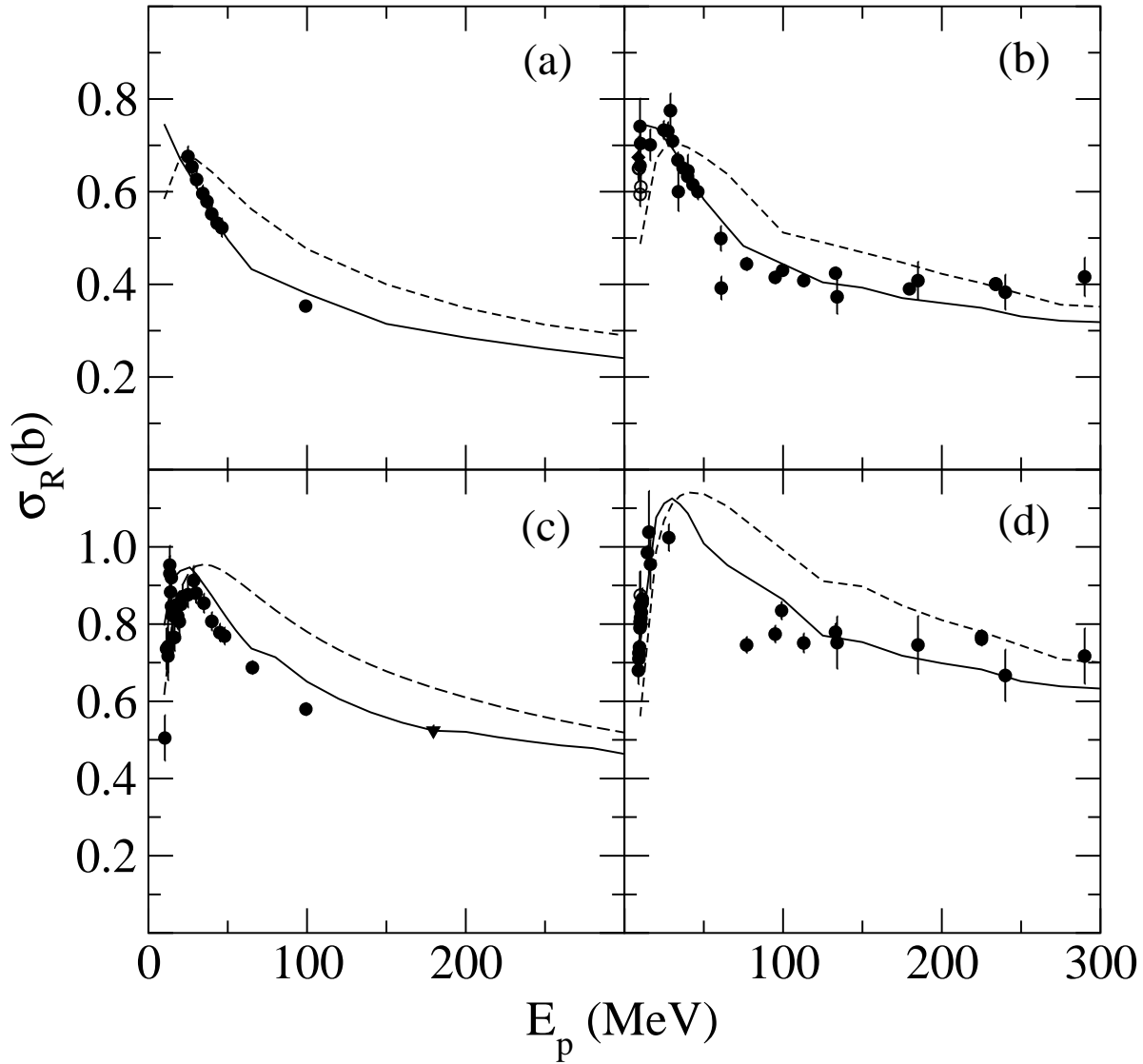}}
\caption{\label{react-F-Al-Ca-Cu}Energy dependencies of $\sigma_R$ for
proton scattering from (a) $^{19}$F, (b) $^{27}$Al, (c) $^{40}$Ca, and
(d) $^{63}$Cu.  The notation is as for Fig.~\ref{react-Li-Be-C-O}.}
\end{figure}

Predictions for proton scattering from $^{19}$F, $^{27}$Al, $^{40}$Ca,
and from $^{63}$Cu respectively are compared with the data in the
segments (a), (b), (c) and (d) of Fig.~\ref{react-F-Al-Ca-Cu}.
(As noted above the wave functions for $^{40}$Ca were obtained from an
SHF model \cite{Ka02}.)  Again all data are well reproduced by the
$g$-folding calculations.  With $^{27}$Al however, three data points
between 180 and 300 MeV are in better agreement with the results of
$t$-folding calculations while one data point, at 61~MeV \cite{Me60},
is in disagreement with both calculations.

Data are shown in this figure again from 10~MeV but with $^{40}$Ca in
particular, the folding model approach is not expected to be reliable
at the energies in the range 10 to 20~MeV. Those excitations energies
correspond to a region of low level density in $^{40}$Ca.  Indeed the
reaction data from $^{40}$Ca show rather sharp resonance-like features
below 20~MeV.  For $^{63}$Cu however, no such sharp structures are
evident in the reaction cross section data and our prediction with a
$g$-folding potential at 10~MeV gives a value in quite reasonable
agreement with observation.  With both $^{40}$Ca and $^{63}$Cu, the
$g$-folding results are in very good agreement with the data for
energies above 20~MeV.  That is in stark contrast to the $t$-folding
results.  The $t$-folding results underestimate the data below 20~MeV
and overestimate considerably the data above 40~MeV.  In the 20 to
40~MeV zone, both calculations give results in reasonable
agreement. Such trends are evident for most heavy nuclei.

\begin{figure}
\scalebox{0.8}{\includegraphics{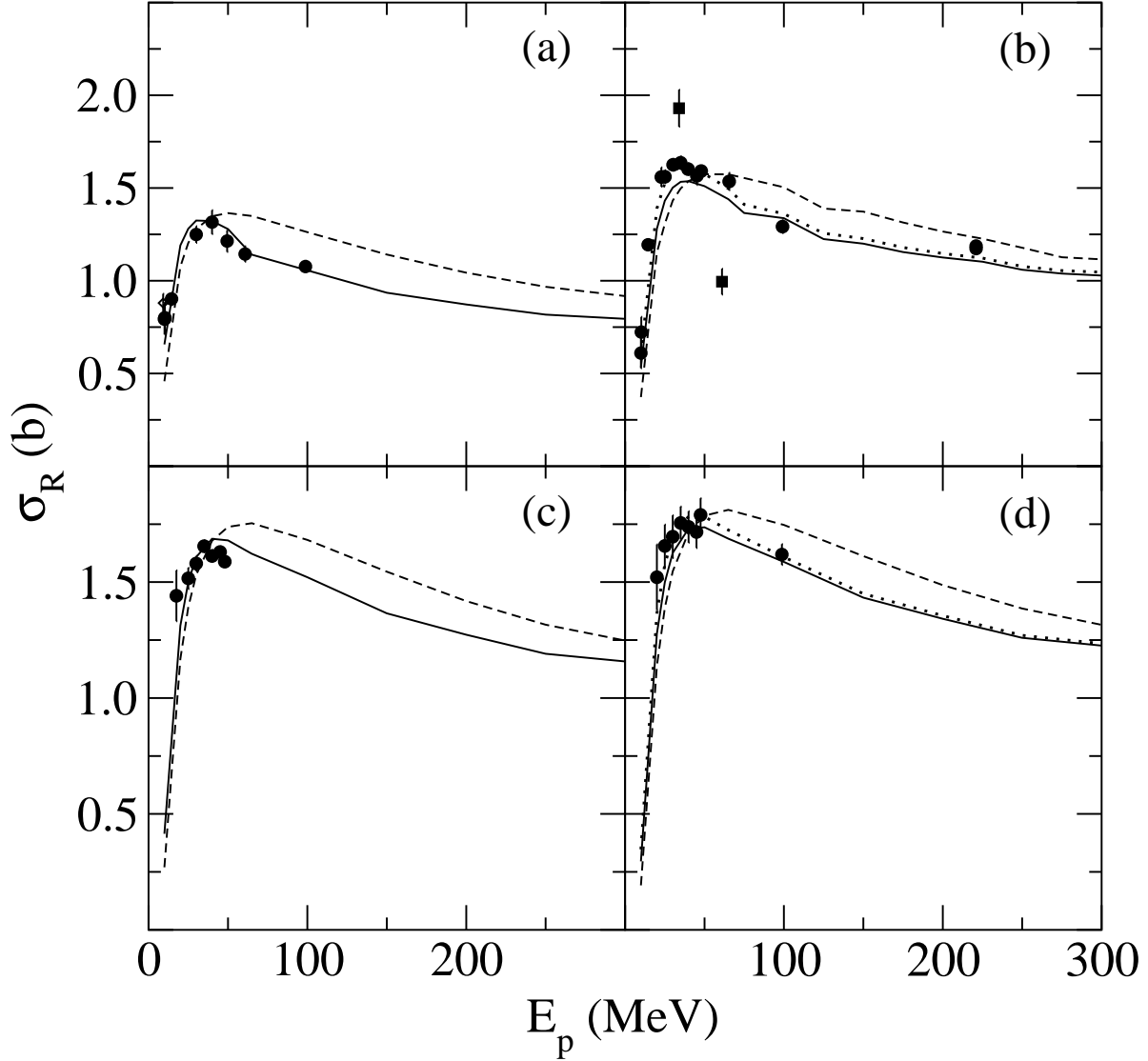}}
\caption{\label{react-zr-sn-ce-tb} Energy dependencies of $\sigma_R$
for proton scattering from (a) $^{90}$Zr, (b) $^{118}$Sn, (c)
$^{140}$Ce, and (d) $^{159}$Tb.  Basic notation is as for
Fig.~\ref{react-Li-Be-C-O}, but now also with dotted curves that
display predictions obtained with the $g$-folding model potentials
formed with the outer most shell neutrons specified by a harmonic
oscillator with an oscillator length 10\% larger.}
\end{figure}

In Fig.~\ref{react-zr-sn-ce-tb}, we present the data and our
predictions of the total reaction cross sections for proton scattering
from $^{90}$Zr, $^{118}$Sn, $^{140}$Ce and $^{159}$Tb.  They are shown
in segments (a), (b), (c), and (d) respectively and compared against
data taken from the relevant references given in Table~\ref{exptab}.
Again results from $g$-folding calculations are in very good agreement
with the data while the $t$-folding results are overestimates at and
above 40~MeV and underestimates the data below 20 MeV.  The third
$p$-$^{118}$Sn total reaction cross section result given in segment
(b) of Fig.~\ref{react-zr-sn-ce-tb} and portrayed in that figure by
the dotted curves, was obtained from a $g$-folding optical potential
formed by varying the surface neutron orbit ($h_{11/2}$) to be that
for an oscillator length increased by 10\% from our basic calculation.
With the (slightly) extended neutron distribution that results, the
$g$-folding potential total reaction cross sections then are in very
good agreement with the data; save for the ubiquitous 61~MeV value.
Likewise there is a datum at 32~MeV at odds with our results.  But
that point also is at odds with other data.  Also in the mismatch
collection, the datum at 17.5~MeV $p$-$^{140}$Ce scattering is
underestimated.  For $^{159}$Tb, the $g$-folding result (solid curve)
is still a quite good replication of data but the calculations
obtained from $g$-folding optical potentials formed by varying the
surface neutron orbit ($h_{9/2}$) to be that for an oscillator length
increased by 10\% (dotted curve) are better.

\begin{figure}
\scalebox{0.8}{\includegraphics{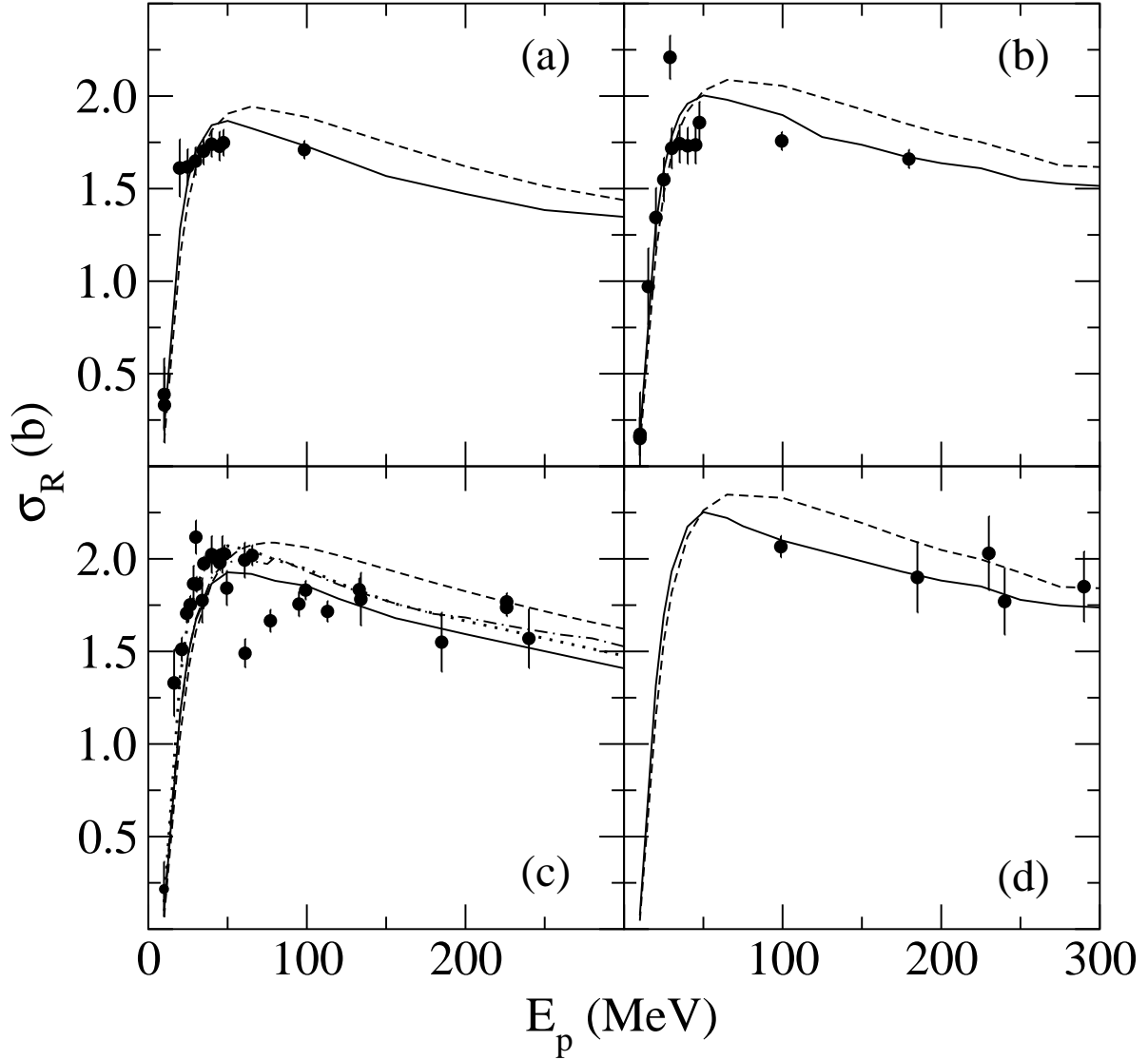}}
\caption[]{\label{react-Ta-Au-Pb-U} Energy dependencies of $\sigma_R$
for proton scattering from (a) $^{181}$Ta, (b) $^{193}$Au, (c)
$^{208}$Pb, and (d) $^{238}$U.  Basic notation is as for
Fig.~\ref{react-Li-Be-C-O}.  For $^{208}$Pb, the dotted curve
is the result of extending the $i_{13/2}$ neutron orbit by increasing
the oscillator length for that shell by 8\%. The dot-dashed curve is
the predictions found using SHF wave functions.}
\end{figure}

In segments (a), (b), (c) and (d) of Fig.~\ref{react-Ta-Au-Pb-U}, we
compare the calculated total reaction cross sections with proton
scattering data from $^{181}$Ta, $^{197}$Au, $^{208}$Pb, and
$^{238}$U.  Again all data are best and well described by the
predictions we make with the $g$-folding model. There is a cross over
regime in energy where the predictions of our $t$- and $g$-folding
potential equate. The differential cross sections do not however and
when such are also considered~\cite{Am00}, preference is given to the
$g$-folding potential results at all energies.

The dotted curve in the case of $^{208}$Pb [segment (c)] results when
the oscillator length for the outer neutron shell ($i_{13/2}$) of the
simple packed shell model we have used to describe the nucleus is
increased by 8\%.  The associated increase in the matter profile
brings the predicted reaction cross sections then in very good
agreement with observation.  Using the SHF wave functions \cite{Br00},
gives the result displayed by the dot-dashed curve in this figure.
Clearly using these new functions has made a slight change to the
predictions found with the simple HO packed model (solid curve); the
modified HO model result is in better agreement with that of the SHF
model.

Exceptional data points are found at 19.8~MeV in $^{181}$Ta where that
data point is underestimated with our calculation by 20\%, at 29~MeV
in $^{197}$Au, and in $^{208}$Pb near 30, 61 and 77~MeV.  Such
exceptional points were also noted in the data from lighter mass
nuclei.  But in most cases, those exceptional point values do not
agree with other measurements made at close values of energy.  For
example other data from $^{208}$Pb taken at 60.8~MeV \cite{Me71} and
at 65.5~MeV \cite{In99} give different results and in fact reaction
cross section values that are consistent with our predictions.  We
note that Menet \textit{et al.} \cite{Me71} argue for a much larger
systematic error in the studies reported in the relevant earlier
experiments.

\subsection{Mass variation of proton total reaction cross sections}
\begin{figure}
\scalebox{0.8}{\includegraphics{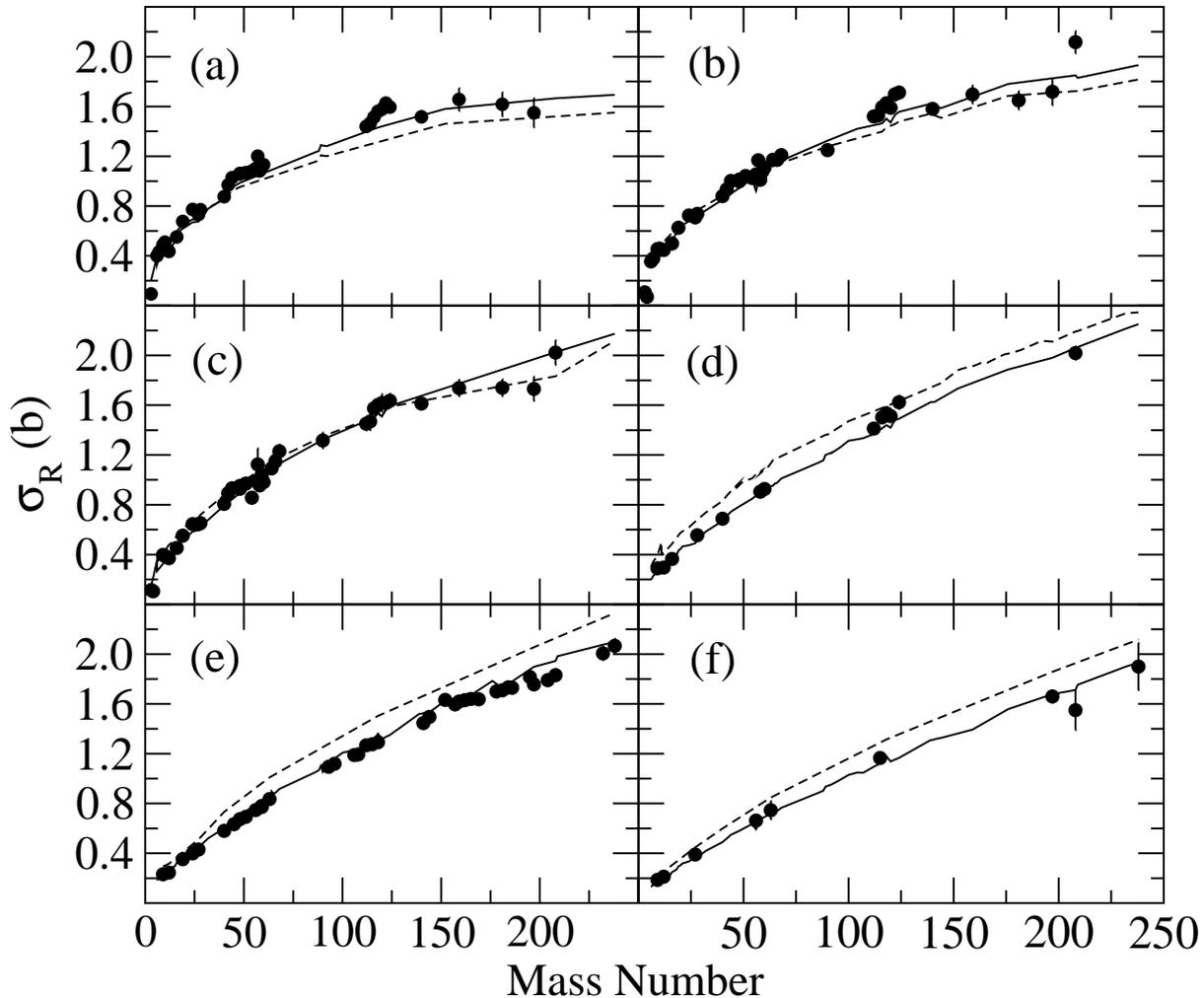}}
\caption{\label{mass-sigma} The mass variations of proton reaction
cross sections at diverse energies. Scattering at energies of 25, 30,
40, 65, 100, and 175~MeV are presented in segments (a), (b), (c), (d),
(e), and (f), respectively.}
\end{figure}

The mass variations of total reaction cross sections for the
scattering of 25, 30, 40, 65, 100 and 175~MeV protons are shown in the
different segments (as labeled) of Fig.~\ref{mass-sigma}.  From that
figure, it is evident that the $g$-folding results are in quite good
agreement with data while the $t$-folding results underestimate most
of the 25 MeV data, are in reasonable agreement with the 30 and 40 MeV
data but overestimate most of the 65 MeV data.  Note that the extended
matter SP states have been used with the $^{208}$Pb and the Sn
isotopes calculations.

The disparities between the $t$- and $g$-folding potential results for
the reaction cross sections are more evident at higher energies.  In
segments (e) and (f) of Fig.~\ref{mass-sigma} we display the mass
variation of the total reaction cross sections measured \cite{Ca96} at
100 and 175~MeV.  Again, the $g$-folding model predictions are in
excellent agreement with the measured values, while the $t$-folding
results overestimate observations typically by 150~mb.  At 100~MeV
proton scattering, proton total reaction cross sections from many
nuclei in the mass range to $^{238}$U have been measured and it is
very clear that the $g$-folding model predictions are in good
agreement with them.  Fewer measurements have been made at 175~MeV,
but they too span the mass range to $^{238}$U and the results of those
measurements also are in very good agreement with the $g$-folding
optical model predictions.

\begin{figure}
\scalebox{0.8}{\includegraphics{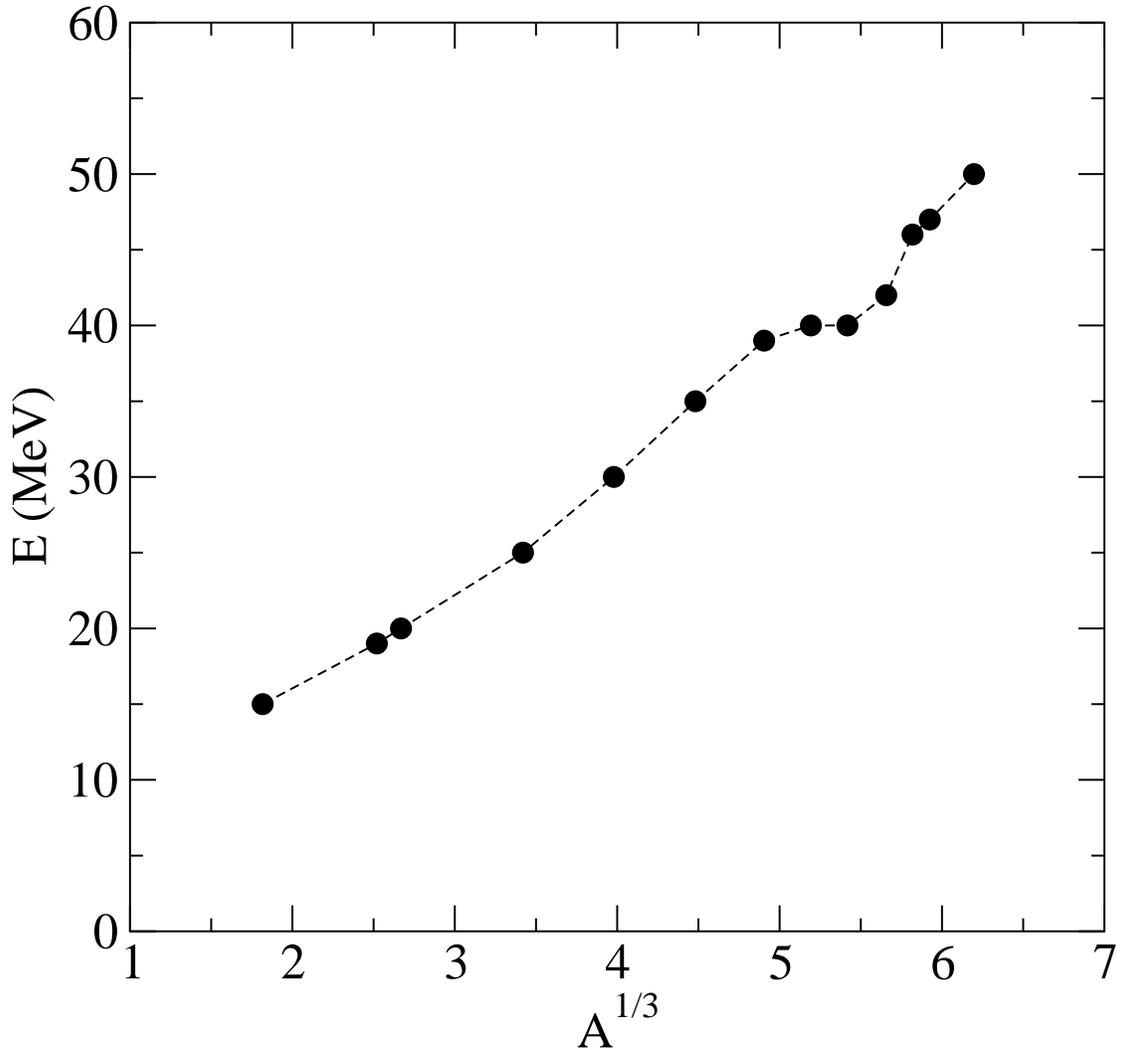}}
\caption{\label{cross-tvsg} The mass-energy equivalence points for $t$-
and $g$-folding results for proton total reaction cross sections. The
line is drawn only to guide the eyes.}
\end{figure}

We have seen in most of the previous figures that, at some energy
point, the total reaction cross sections obtained by the $t$-folding
and $g$-folding calculations are same. In Fig.~\ref{cross-tvsg} we
display those crossing energy values as a function of $A^{1/3}$; the
energy is in MeV. It clearly indicates that there is a line in mass
and energy above and below which the $g$-folding calculation results
are smaller and larger respectively than the corresponding $t$-folding
ones.

\subsection{Energy/mass variations of neutron total cross sections}

Accurate measurements of neutron total reaction cross sections are far
more difficult to achieve than their proton counterparts. Indeed
usually those cross sections are obtained by subtracting the
elastic from the total scattering cross section. While both of those
cross sections can be measured with some accuracy, the subtraction of
two large numbers with attendant uncertainties is subject to numerical
problems. We note, however, that a new technique from Japan for
measuring neutron total reaction cross sections utilizing in-beam and
out-beam methods similar to those used in proton scattering shows
promise.  Nevertheless, herein, we concentrate on analyses of total
scattering cross section data.

The data we have chosen to analyze have been taken from a recent
survey by Abfalterer {\em et al.}~\cite{Ab01}. That survey includes
data measured at LANSCE that are supplementary and additional to those
published earlier by Finlay {\em et al.} \cite{Fi93}. From that recent
data compilation, we have selected cases for study having target
masses that span the stable mass range and which includes as many of
the nuclei as possible for which we have analyzed proton scattering
data.  Specifically we calculate neutron total scattering cross
sections for $^6$Li, $^{12}$C, $^{19}$F, $^{40}$Ca, $^{89}$Y,
$^{184}$W, $^{197}$Au, $^{208}$Pb and $^{238}$U. In some cases the
data taken were from natural targets; we indicate those as discussion
of them occurs in the text.  Given that analyses with the $g$-folding
optical potential has been found the more appropriate with the proton
scattering analyses, we only present in the ensuing figures, results
obtained from the $g$-folding optical models for neutron
scattering. In forming those neutron optical potentials, the structure
models used were those that defined proton scattering.  The effective
$NN$ interactions were as well.

\begin{figure}
\scalebox{0.8}{\includegraphics{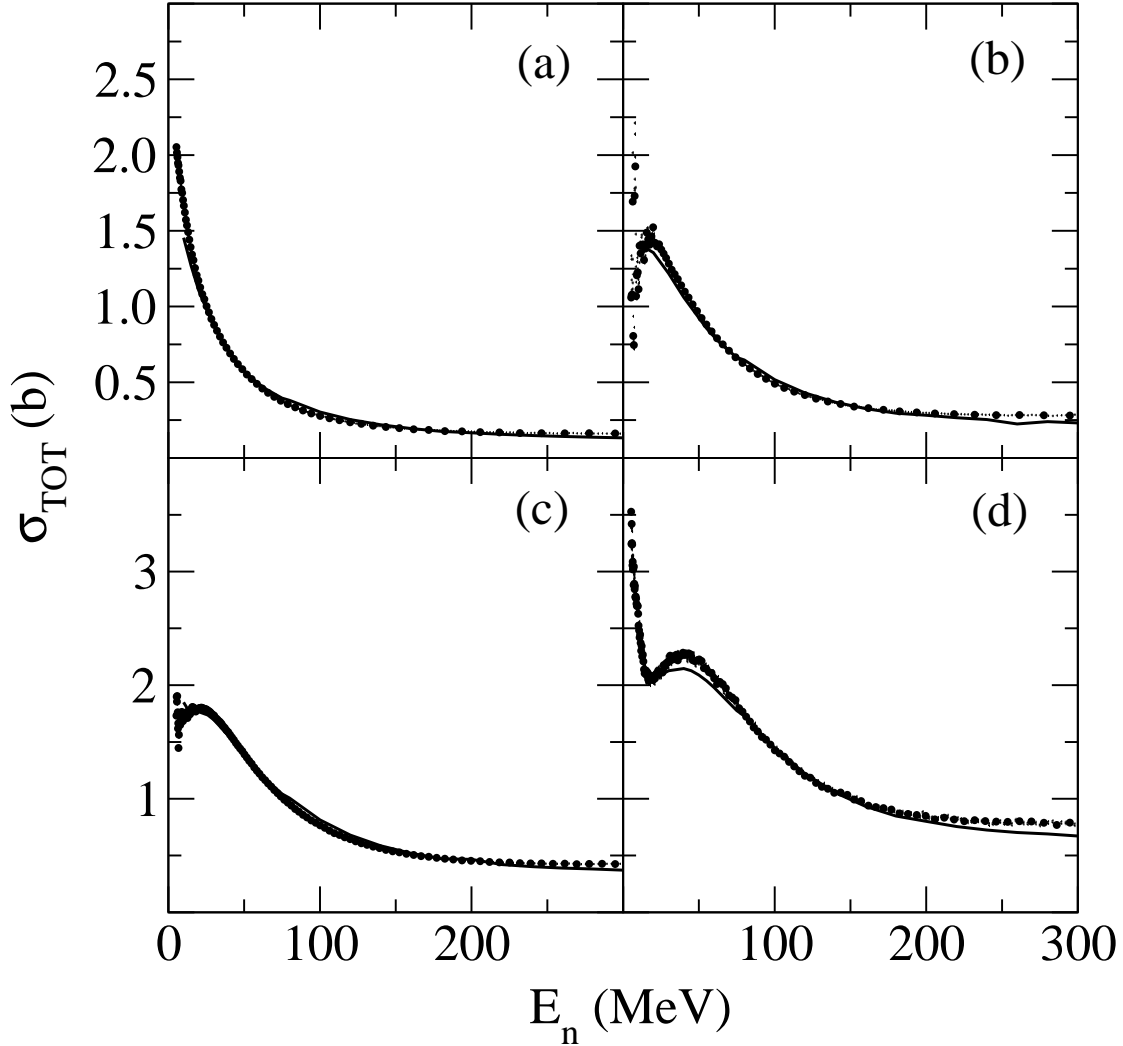}}
\caption[]{Total cross sections for neutrons scattered from (a)
$^6$Li, (b) $^{12}$C, (c) $^{19}$F, and (d) $^{40}$Ca. The data are
those of Abfalterer {\em et al.} \cite{Ab01}; in (b) the data
correspond to scattering from natural carbon.}
\label{neutrons1}
\end{figure}

Predictions of the total cross sections for neutrons scattered from
$^6$Li, $^{12}$C, $^{19}$F, and $^{40}$Ca are compared to the data in
Fig.~\ref{neutrons1}.
The models for $^6$Li, $^{12}$C, and $^{40}$Ca to specify the
densities were those used in the calculations of proton
scattering. The model for $^{19}$F was a $0\hbar\omega$ shell model
calculation using the WBT interaction of Warburton and Brown
\cite{Wa92}. The oscillator parameter for the HO SP functions in that
case was 1.855~fm, chosen to reproduce the rms radius of $^{19}$F
(2.9~fm \cite{Vr87}). In all cases except for $^{40}$Ca, there is
excellent agreement with the data, and the turnover in the cross
sections at $\sim 20$~MeV for natural carbon and $^{19}$F are
predicted. The low-energy structure in the cross section for $^{40}$Ca
is reproduced, although the peak at 40~MeV is underpredicted by
10\%. Also for $^{40}$Ca the cross section at 300~MeV is
underpredicted.

\begin{figure}
\scalebox{0.8}{\includegraphics{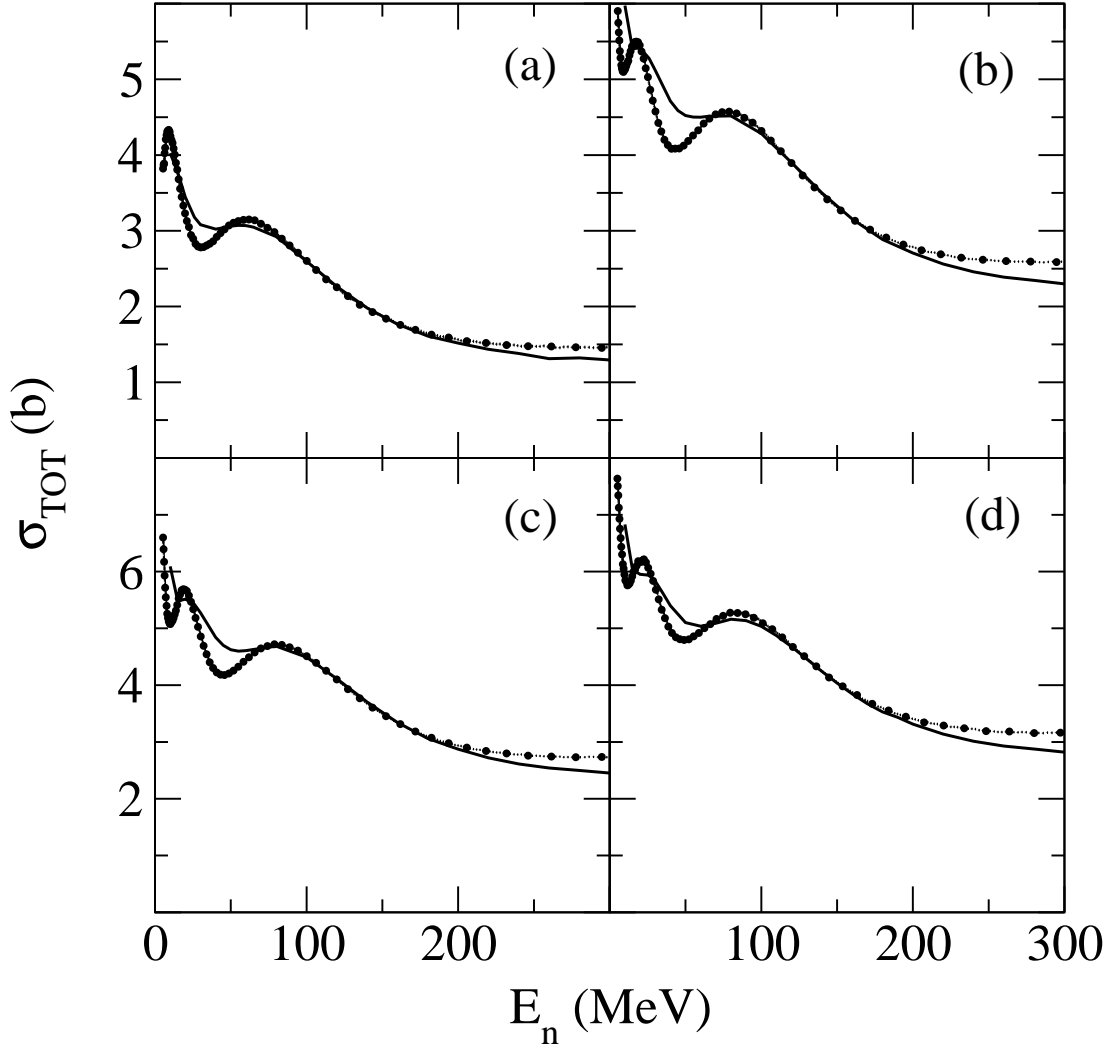}}
\caption[]{Total cross sections for neutrons scattered from (a)
$^{89}$Y, (b) $^{184}$W, (c) $^{197}$Au, and (d) $^{238}$U. The data
are those of Abfalterer {\em et al.} \cite{Ab01}; in (b) the data
correspond to scattering from natural tungsten.}
\label{neutrons2}
\end{figure}

Our predictions of the total cross sections for neutrons scattering
from $^{89}$Y, $^{184}$W, $^{197}$Au, and $^{238}$U are compared to
the data in Fig.~\ref{neutrons2}.
In the case of $^{184}$W, the data were taken from scattering using
natural tungsten as the target. The structure of three of these nuclei
was obtained from a simple packed model as used previously. For
$^{89}$Y, however, the shell model of Ji and Wildenthal \cite{Ji89}
was used to specify the density. The four results exhibit the same
features. The predicted cross sections agree well with the
experimental results between 70 and 200~MeV, with the peak cross
section values slightly underpredicted. Below 70~MeV the cross
sections are overpredicted but the structural character exhibited in
the data is found. To obtain better agreement at energies 40 to 70~MeV
at least, improvements in the SP functions are required. Improvements
such as by using a SHF model seem needed, given the results we shall
present next for $^{208}$Pb.  Above 200~MeV, all results underpredict
the data. We believe that this indicates that our present effective
interaction for those energies at and about the pion production
threshold need be improved.  Possibly more explicit contributions from
the $\Delta$ resonance in the $NN$ interaction in defining the
effective $NN$ force to be used in the $g$-folding giving the optical
potentials are needed~\cite{Fu01}.

\begin{figure}
\scalebox{0.8}{\includegraphics{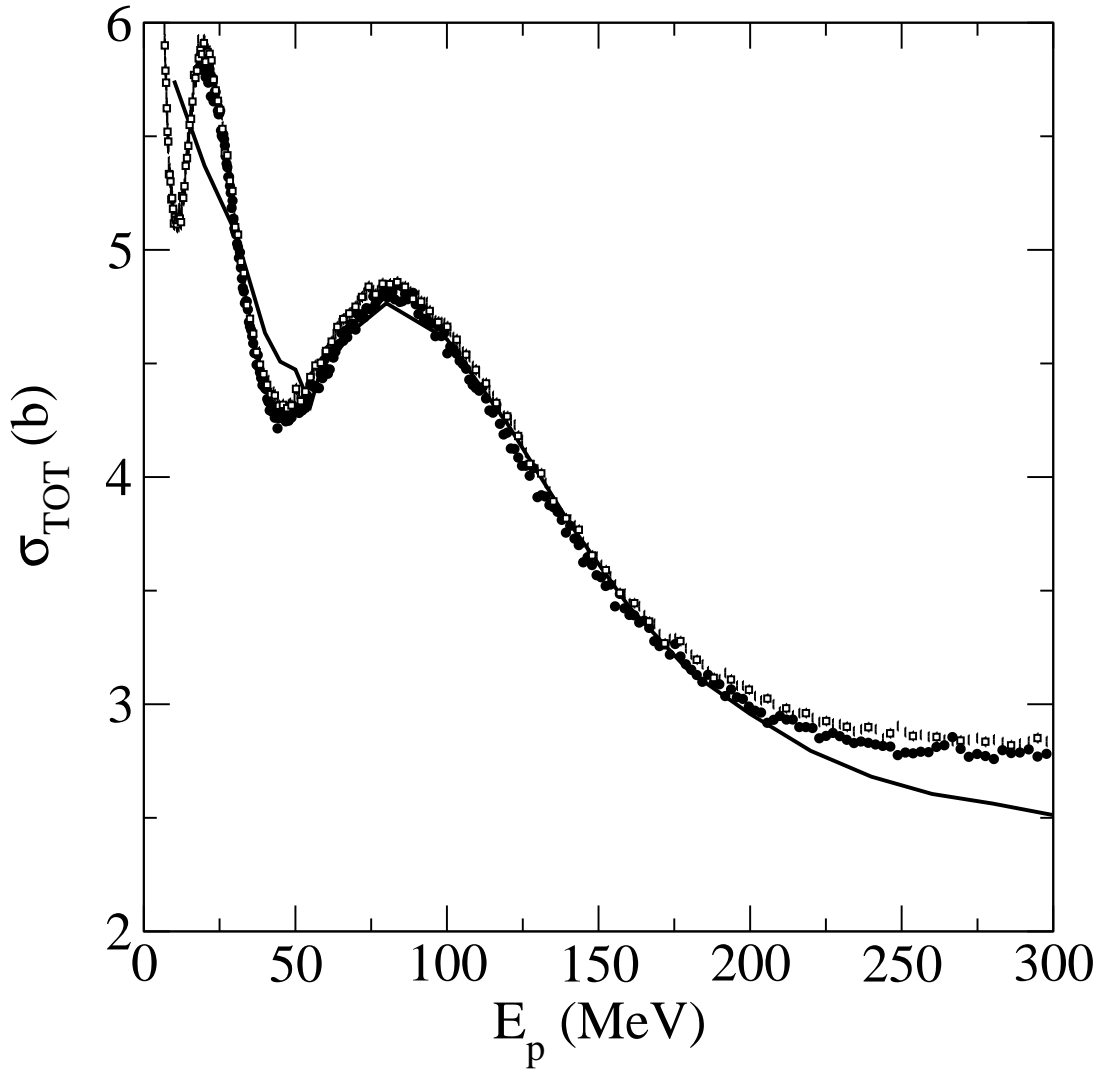}}
\caption[]{Total cross section for the scattering of neutrons from
$^{208}$Pb. The results of our calculations are compared to the data
of Finlay {\em et al.} \cite{Fi93} (circles) and of Abfalterer {\em et
al.} \cite{Ab01} (squares). The latter set correspond to scattering
from natural Pb.}
\label{neutpb}
\end{figure}

As stated above in the case of proton scattering, $^{208}$Pb is a
special case. To specify the density of $^{208}$Pb we have used the
SHF calculation of Brown \cite{Br00,Ka02}. The results of our
calculations of the total neutron scattering cross section from
$^{208}$Pb using that model are given in Fig.~\ref{neutpb}.
Therein, we compare our prediction with the data of Finlay {\em et
al.}  \cite{Fi93} and of Abfalterer {\em et al.} \cite{Ab01}. The
latter set correspond to scattering from natural Pb. From 60
to 200~MeV, the agreement with the data is excellent, as it was for
the other nuclei also. Below 60~MeV, the energy dependence is
reproduced, although the minimum at 50~MeV is slightly
overpredicted. Note that when the oscillator model is used, the
calculated results do not reproduce the data in this energy regime as
adequately \cite{De01}.  Above 200~MeV, the predicted cross section
falls too sharply as it did for the other scattering cases. The
variation of the structure model from HO to SHF did not influence much
the result above 200~MeV.  Again we think that our effective force may
be at fault for this data at these energies.  The mismatch above
200~MeV is influenced by the mass of the target however, with little
problem evident in the light mass results as shown in
Fig.~\ref{neutrons1}.  Given that those light mass nuclei as
characterized as mostly ``surface'', we also believe that the problems
with the effective interactions at the higher energies relates to the
character of the force at central densities.

\section{Conclusions}

A microscopic model of the nucleon-nucleus optical potential in
coordinate space has been used to predict successfully the total
reaction cross sections of nucleons from nuclei.  That optical
potential has been formed by folding complex energy- and
density-dependent effective $NN$ interactions with OBDME of the target
obtained primarily from shell models of the nuclei.  As the approach
accounts for the exchange terms in the scattering process, the
resulting complex and energy dependent optical potential also is
non-local.  We have found that it is crucial to use effective $NN$
interactions which are based upon realistic free $NN$ interactions and
which allow for modification from that free $NN$ scattering form due
to nuclear medium effects of Pauli blocking and an average mean field.
For optimum results, and for the light masses in particular, it is
essential also to use the best (nucleon based) model specification of
nuclear structure available.  Marked improvement in results were
obtained when, for $^9$Be and $^{12}$C in this study, complete
$(0+2)\hbar\omega$ shell model calculations were used to define the
OBDME required in the folding processes. For scattering from heavy
nuclei at around 300~MeV, our results indicate the need for
improvement in the effective $NN$ force possibly by explicit inclusion
of $\Delta$ effects but also of those associated with the interaction
at central field densities.

\begin{acknowledgments}
This research was supported by a research grant from the Australian
Research Council, and also by DOE contract no. W-7-405-ENG-36.
\end{acknowledgments}

\bibliography{Reaction-fullpaper}

\end{document}